\documentclass[amsmath,amssymb,pra,aps,showpacs,twocolumn,10pt,superscriptaddress,tightenlines]{revtex4-2}
\usepackage[colorlinks=true,citecolor=blue,linkcolor=red,urlcolor=blue]{hyperref}
\usepackage[dvipdfmx]{graphicx}
\usepackage{dcolumn}
\usepackage{bm}
\usepackage[utf8]{inputenc}
\usepackage[T1]{fontenc}
\usepackage{lmodern}
\usepackage{cleveref}
\usepackage{color}
\usepackage{amsmath}
\usepackage{amssymb}
\usepackage{amsfonts}
\usepackage{latexsym}
\usepackage{multirow}
\usepackage{booktabs}
\usepackage{comment}
\usepackage{ulem}

\crefname{equation}{}{}

\begin{document}

\title{Beyond Optimization: Harnessing Quantum Annealer Dynamics for Machine Learning}

\author{Akitada Sakurai}\email{Akitada.Sakurai@oist.jp}
\affiliation{Okinawa Institute of Science and Technology Graduate University, Onna-son, Okinawa 904-0495, Japan}

\author{Aoi Hayashi}
\affiliation{Okinawa Institute of Science and Technology Graduate University, Onna-son, Okinawa 904-0495, Japan}

\author{Tadayoshi Matsumori}
\affiliation{Toyota Central R\&D Labs., Inc., Yokomichi 41-1, Nagakute, Aichi, Japan}

\author{Daisuke Kaji}
\affiliation{DENSO CORPORATION, 1-8-15, Kounan, Minato-ku, Tokyo 108-0075, Japan}

\author{Tadashi Kadowaki}
\affiliation{Global Research and Development Center for Business by Quantum-AI technology (G-QuAT), National Institute of Advanced Industrial Science and Technology (AIST),}
\affiliation{DENSO CORPORATION, 1-8-15, Kounan, Minato-ku, Tokyo 108-0075, Japan}

\author{Kae Nemoto}\email{kae.nemoto@oist.jp}
\affiliation{Okinawa Institute of Science and Technology Graduate University, Onna-son, Okinawa 904-0495, Japan}

\date{\today}
\begin{abstract}
Quantum annealing is typically regarded as a tool for combinatorial optimization, but its coherent dynamics also offer potential for machine learning. 
We present a model that encodes classical data into an Ising Hamiltonian, evolves it on a quantum annealer, and uses the resulting probability distributions as feature maps for classification.  
{Experiments on the quantum annealer machine with the Digits dataset, together with simulations on MNIST, demonstrate that short annealing times yield higher classification accuracy, while longer times reduce accuracy but lower sampling costs. } 
We introduce the participation ratio as a measure of the effective model size and show its strong correlation with generalization.  
\end{abstract}

\maketitle

The quantum annealer (QA) was originally conceived as a heuristic framework for tackling combinatorial optimization problems by exploiting quantum fluctuations~\cite{Kadowaki1998,DasArnab2005,Santoro2006,DasArnab2008,Morita2008,Albash2018,Hauke2020}. The basic principle is to embed an optimization problem into an Ising Hamiltonian, initialize the system in a simple ground state, and then slowly evolve the Hamiltonian so the system remains in or near the ground state.  Over the last decade, this approach has motivated the development of large-scale QA hardware, most prominently the D-Wave systems currently feature thousands of programmable qubits \cite{mcgeoch2022adiabatic,Dwave}. However, the same coherent processes that enable ground-state search also constitute a controllable quantum dynamical resource.

Since QA can be regarded as a specialized form of adiabatic quantum computation~\cite{DasArnab2005,Santoro2006,DasArnab2008,Morita2008,Albash2018,Hauke2020}, its scope has steadily broadened~\cite{Wurtz:2022aa}. Recent studies have investigated its role in simulating nonequilibrium quantum dynamics~\cite{King2023,King2025}, probing quantum phase transitions, and supporting machine learning tasks~\cite{neven2009,Mott2017aa}. In particular, annealers have been applied as Boltzmann samplers~\cite{Crawford2018,Liu2018,Sato:2021,Shibukawa2024}, as approximate solvers for combinatorial formulations of machine learning problems~\cite{Li2018,Salloum2025}, and more recently as random feature generator for machine learning model~\cite{Noori2020}. These developments reflect a shift in perspective: QAs are increasingly recognized not only as heuristic optimizers but also as programmable dynamical systems capable of processing information in ways not easily reproduced by classical algorithms.

In this Letter, we build on this perspective by proposing a machine learning framework that directly exploits the coherent dynamics of a QA as a source of expressive quantum features~\cite{Fujii2017,Fujii2018,Mitarai2018,Fujii2020,Nakajima2019,Suzuki2022,Bravo2022, Pfeffer2022,Mujal2023,Hayashi2023}.  Instead of restricting the device to optimization or sampling roles, we encode classical input data into the Ising Hamiltonian of the annealer, allow the system to evolve under the QA dynamics, and use the resulting output probability distributions as feature vectors for classification. This viewpoint highlights the annealer as a quantum feature generator, expanding its role beyond optimization and sampling.

The significance of this approach is threefold.  First, we demonstrate its feasibility on real hardware by performing $8\times 8$-handwritten digit classification~\cite{scikit-learn,DIGIT1998}  using the D-Wave Advantage System 7.1~\cite{boothby2020,Dwave2025}  (hereinafter referred to as QA machine), confirming that meaningful features can be extracted under realistic experimental conditions.  Second, we extend the method to the MNIST dataset~\cite{Deng2012}, revealing a systematic dependence of model performance on the annealing time:
short-time dynamics yield superior generalization, while longer times degrade accuracy but reduce the number of required measurements (shots). 
Third, we introduce the participation ratio (PR)~\cite{Edwards1972aa} and its variants as quantitative indicators of the model’s effective size and show that they strongly correlate with classification accuracy, providing a principled diagnostic for model capacity and generalization.

\begin{figure*}[tb]
\centering
\includegraphics[width=0.93\textwidth]{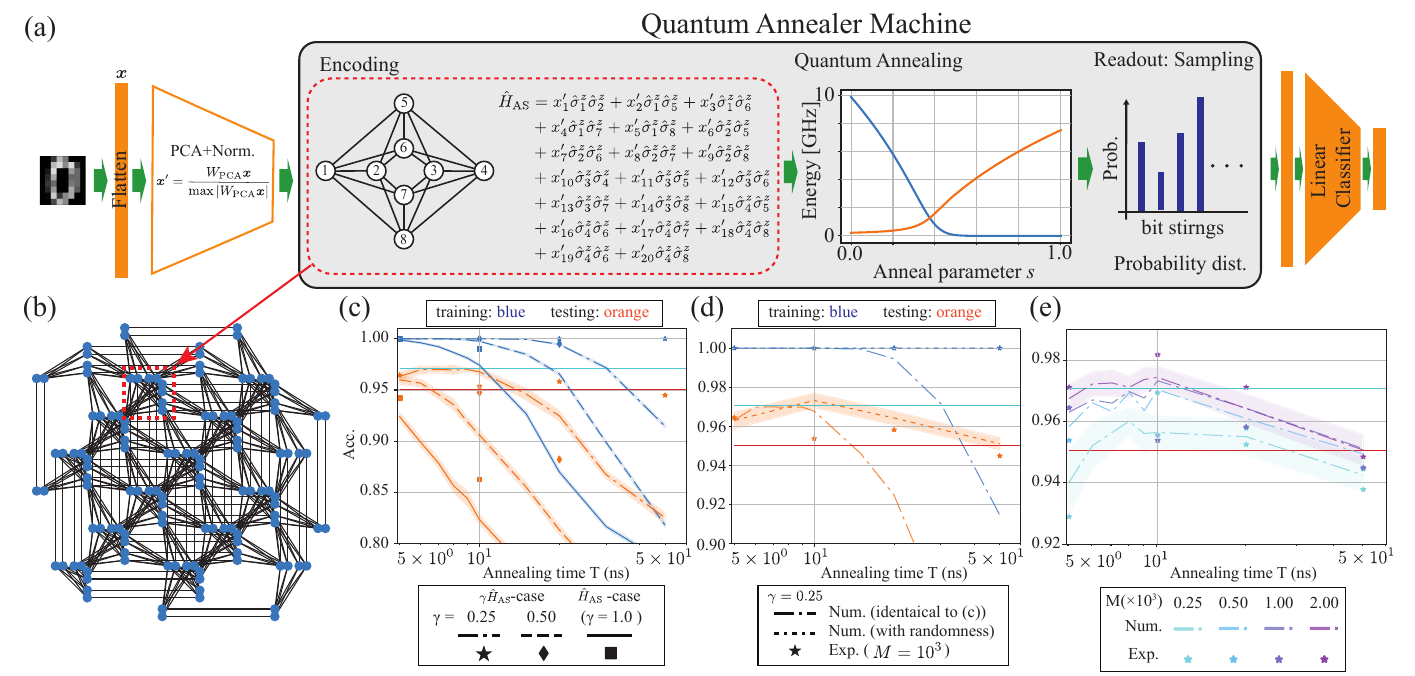}
 \caption{
 \textbf{Overview of the QA-based machine learning model and experimental results.}  (a) Schematic of our quantum machine learning (QML) model implemented on a QA, e.g., the D-Wave Advantage System 7.1. Here, the PCA-based dimensionality compression is expressed as a linear transformation using the matrix $W_\mathrm{PCA}$, and assign each element of this 20-dimensional vector $\boldsymbol{x}'$ to the coupling constants of the final Hamiltonian $\hat{H}_\mathrm{AS}$ of the QA. (b)Structure of the Pegasus graph and a representative subgraph used.  (c) Classification accuracy as a function of annealing time for different values of the hyperparameter $\gamma$.  We collected 1,000 shots for each image and performed the learning procedure ten times. The solid, dashed, and dash-dotted curves correspond to $\hat{H}_\mathrm{AS}$ and $\gamma\hat{H}_\mathrm{AS}$ cases with $\gamma = 0.25$, and $0.5$, respectively.  Square, diamond, and star markers denote experimentally obtained accuracy rates for $\gamma = 0.25, 0.5$.  (d) Accuracy rates for $\gamma = 0.25$ in the presence of Hamiltonian randomness. For each image, the output probability distribution was averaged over 100 random realizations of the Hamiltonian.  Using these averaged probabilities, we simulated measurements with 1,000 shots.  The blue and orange dash-dotted lines show the averaged training and testing accuracy, respectively, each averaged over 10 independent measurement simulations.
The colored areas indicate the corresponding standard deviations. (e) Testing accuracy versus number of shots for simulation and experiment.   Testing accuracy rates obtained from both numerical simulations and experiments are plotted as a function of the number of measurement shots $M$.  Star-shaped points denote the experimentally measured testing accuracy. Dash-dotted curves correspond to the accuracy obtained from numerical simulations including Hamiltonian randomness. The curves represent averages while the shaded areas indicate the associated standard deviations, shown here explicitly only for $M = 250$ and $M = 2000$. In (e)-(d), the red solid line shows the peforamnce of the linear model (without QA).
 }
\label{DwaveAd71}
\end{figure*}

We begin by exploring the feasibility of our quantum machine learning (QML) model implemented on the QA machine. The QA performs the time evolution of the transverse-field Ising Hamiltonian $ \hat{H}_\mathrm{Ising}(s) = -\frac{A(s)}{2} \hat{H}_1 + \frac{B(s)}{2} \hat{H}_2,$ where $\hat{H}_1= \sum_{l=1}^N \sigma_l^x$ is the initial transverse magnetic field Hamiltonian, and $ \hat{H}_2 = \sum_{l > m}^N J_{lm} \sigma_l^z \sigma_m^z + \sum_{l=1}^N h_l \sigma_l^z$  the final problem Hamiltonian.  The coefficients $J_{lm}$ and $h_l$ represent the two-qubit coupling strengths and local longitudinal magnetic fields, respectively. The annealing schedule is defined by the functions $A(s)$ and $B(s)$. To achieve quantum annealing within a short, coherence-preserving timescale, we used the Fast Anneal mode provided by D-Wave~\cite{mehta2025,Dwave2025}.  This restricts the use of the longitudinal magnetic field ($h_l$) and allows only the two-body interaction terms ($J_{lm}$).  
The coupling constants are constrained to the range $J_{lm} \in [-1,1]$.  
In our numerical calculations, the functional forms of $A(s)$ and $B(s)$ in Fast Anneal mode are based on data from Ref.~\cite{Dwave2025}.

The QA thus realizes the time evolution underlying our QML model.  
The annealing process starts from the initial state $ |\phi(0)\rangle = \prod_{l=1}^N \frac{|0\rangle_l + |1\rangle_l}{\sqrt{2}}$, where $|0\rangle_l$ and $|1\rangle_l$ are eigenstates of $\hat{\sigma}_l^z$ for each qubit.  
The state at time $T$ is given by the time-dependent Hamiltonian $\hat{H}(t)$ as $|\phi(T)\rangle = \mathcal{T} \exp\left( -i  \int_0^T \hat{H}(t, \boldsymbol{x})\, dt  /\hbar\right) |\phi(0)\rangle$, where $\mathcal{T}$ denotes the time ordered product.

This evolution encodes the input parameters $\boldsymbol{x}$ into the system’s quantum state, forming the basis of our feature-mapping procedure. By measuring the final state $|\phi(T)\rangle$ in the computational basis, we obtain the output probability distribution $\boldsymbol{P}(\boldsymbol{x})$ corresponding to each bit string: 
\begin{equation}\label{eq:prob}
\boldsymbol{ P}(\boldsymbol{x}) =
 \left( |\langle 00\cdots 0 |\phi\rangle|^2,\, \cdots,\, |\langle 11\cdots 1 |\phi\rangle|^2
 \right)^\top.
\end{equation}
This distribution is estimated by performing multiple measurements (shots), and the resulting probability vector is treated as a classical feature representation of the input $\boldsymbol{x}$. To ensure comparability across samples, we standardize the obtained output distributions based on the training data (known as the StandardScaler of \texttt{scikit-learn}~\cite{scikit-learn}), applying the same transformation to the test data.  This vector is denoted by $\boldsymbol{u}(\boldsymbol{x})$.

Our classifier is a simple multi-class perceptron \cite{haykin2009}, whose output is defined as $\boldsymbol{f}_{\boldsymbol{W},\boldsymbol{b}}(\boldsymbol{x}) = \mathrm{Softmax}\!\left( \boldsymbol{W} \cdot \boldsymbol{u}(\boldsymbol{x}) + \boldsymbol{b} \right)$, where $\boldsymbol{W}$ is a $(c \times 2^N)$-dimensional weight matrix and $\boldsymbol{b}$ is a c-dimensional bias vector, where $c$ is the total number of classes of the given task. A cross-entropy loss function is used for learning, where $(\boldsymbol{x}, \boldsymbol{t})$ denotes each training example and its target label~\cite{Akitada2022,Sakurai25Opt}. Optimization is performed using a combination of the AdaGrad and mini-batch methods to ensure stable convergence~\cite{Mach2011}.

For the experimental demonstration, we implemented the QML model on the QA machine, shown schematically in Fig.~\ref{DwaveAd71}(a). This device features 5,554 qubits connected in a topology known as the Pegasus graph.  We use a subgraph of eight qubits (see Fig.~\ref{DwaveAd71}(a)).   As a benchmark dataset, we employed the Digits dataset provided by \texttt{scikit-learn}, consisting of $8\times8$ grayscale images of handwritten digits. There are 1,797 images categorized into 10 classes (0–9).  Of these, 1,347 images are used for training and 450 for testing. We apply principal component analysis (PCA) to compress each image representation into a 20-dimensional feature vector.  Each component of this reduced vector is then encoded into the coupling constants of the final Ising Hamiltonian, $\hat{H}_\mathrm{AS}$, of the QA, see Fig.~\ref{DwaveAd71} (b).

In the experiment, the annealing time $T$ plays a crucial role in determining the system's performance, see Fig.~\ref{DwaveAd71} (c). Due to experimental constraints, the minimum accessible annealing time is approximately $5~\text{ns}$, corresponding to the shortest duration available under the fast-anneal protocol.   Although the training accuracy reaches relatively high values, the testing accuracy within this annealing-time regime remains below that achieved by a classical linear classifier. Extending the annealing time generally results in a decrease in accuracy.

These observations suggest that exploring a shorter annealing-time regime, $T < 5~\text{ns}$, is essential for improved performance~\cite{Dwave2025}.
To effectively emulate shorter annealing times within the hardware’s operational limits, we rescale the energy scale of the Hamiltonian using a parameter $\gamma \in [0, 1]$, as $\hat{H}_\text{AS}\rightarrow \gamma \hat{H}_\text{AS}$. This parameter $\gamma$ can be experimentally controlled by adjusting the normalization of the encoded data as $\boldsymbol{x}' \rightarrow \gamma \boldsymbol{x}'$.   For $\gamma < 1$, the overall energy scale of the annealer Hamiltonian is reduced, effectively simulating a faster annealing process.

In Fig.~\ref{DwaveAd71} (c), we plot the numerically obtained accuracy rates as a function of the annealing time for various values of $\gamma$ within the short-time regime ($T \le 5~\text{ns}$).  The results indicate that the curves for $\gamma = 0.25,\;0.5$ exhibit similar profiles to that of $\gamma = 1.0$, but shifted along the time axis. Importantly, the testing accuracy exceeds that of the classical linear classifier within a certain annealing-time range, demonstrating the potential advantage of appropriately scaled quantum annealing.

The experimental results of the image classification task performed on the D-Wave Advantage System are presented in Fig.~\ref{DwaveAd71} (c). As shown in Fig.~\ref{DwaveAd71} (d), the testing accuracy surpasses that of the classical linear classifier, demonstrating the capability of the QA to generate effective feature representations even under realistic noise conditions, which is specified later.

Although the hardware noise inevitably affects the measurements, the comparison between experimental and numerical results reveals that the experimentally obtained accuracy is consistently higher than the simulated one.  This behavior suggests an effective extension of the annealing time scale in the experiment compared to the idealized numerical model.  To quantitatively examine this effect, we analyze the deviation between the experimental probability distribution $P(\boldsymbol{x})$ and that obtained from numerical simulation.   Let $P^{\text{exp}}(T^{\text{exp}})$ denote the experimentally measured probability distribution with the annealing time $T^{\text{exp}}$ for a fixed image , and $P^{\text{sim}}(T)$ denote the corresponding numerically simulated distribution for variable $T$.   We define the squared error $L(T) = \sum_{i=1}^{2^N}\left(P^{\text{exp}}_i(T^{\text{exp}}) - P^{\text{sim}}_i(T)\right)^2$,
which quantifies the difference as a function of $T$.  

By averaging $L(T)$ over 100 training images, we find that the value of $T^*$ minimizing the error is systematically shorter than the actual annealing time $T^{\text{exp.}}$ used in the experiments. This indicates that the experimentally obtained probability distributions resemble those of shorter-time numerical evolutions, which explains the observed shift in the accuracy shown in Fig.~\ref{DwaveAd71} (c).

Regardless of the specific origin of noise in the experiment, its presence does not degrade the classification accuracy; on the contrary, it appears to enhance the performance of the QML model.  To investigate this noise-assisted mechanism, we introduce controlled randomness into the Hamiltonian $\hat{H}_\text{AS}$.   In particular, we add random perturbations uniformly sampled from the range $[-0.1, 0.1)$ to the coupling strengths $J_{lm}$ and introduce random longitudinal fields with amplitudes drawn from the same interval, even though explicit longitudinal-field control is unavailable in the fast-anneal protocol. Fig.~\ref{DwaveAd71} (d) shows the resulting accuracy rates for $\gamma = 0.25$ when such randomness is included. 
The behavior of both training and testing accuracies closely matches the experimental results, suggesting that the experimentally observed performance can be partially attributed to similar stochastic effects.  Although this injected randomness does not constitute an exact model of the hardware noise, the agreement indicates that our QML algorithm is intrinsically robust to moderate random fluctuations in the physical Hamiltonian.

Next, we assess the effect of the number of measurement shots on the quality of the probability distribution read out from the QA. Since the distribution does not need to be reconstructed, the goal of this analysis is to determine how few shots are sufficient to achieve reliable classification. Fig.~\ref{DwaveAd71} (e) shows the experimentally obtained testing accuracy for different numbers of shots, along with numerical simulations including the Hamiltonian randomness.  As indicated by the figure, $0.5\times 10^3$ shots per image are sufficient in the experiment to achieve testing accuracy exceeding that of a classical linear classifier.  This number of shots is significantly lower than what previous models have required \cite{Akitada2025,Sakurai25Opt}, highlighting one potential advantage offered by the QA.

Finally, we assess the feasibility of this QML model for larger, more practical datasets. To evaluate this numerically, we extend the model to classify the MNIST dataset which consists of 70,000 $28 \times 28$ pixel handwritten digit images. Following standard practices, 60,000 images are used for training, and 10,000 for testing.

For generality, we consider a simplified QA model given by $\hat{H}_\mathrm{Ising}(t) = \left(1-t/T\right) \hat{H}_1 + \left(t/T\right) \hat{H}_2$, which is commonly employed in quantum annealing theoritical and numerical studies. The initial Hamiltonian is  $\hat{H}_1 = -\hbar \sum_{l=1}^N \sigma_l^x$, and the final Hamiltonian is taken as a nearest-neighbor Ising model:  $ \hat{H}_2 = \hbar \sum_{l =1}^{N-1} J_{l} \sigma_l^z \sigma_{l+1}^z + \hbar \sum_{l=1}^N h_l \sigma_l^z$. Unlike the fast-anneal setting of the D-Wave device, here we assume that longitudinal magnetic fields can be applied freely.
As with the DIGITS dataset, dimensionality reduction using PCA and normalization is performed on the MNIST images.
For the new $\hat{H}_2$, we encode the input $\boldsymbol{x}^\prime$ as follows:
\begin{equation}
\hat{H}_2(\boldsymbol{x}^\prime) = \hbar \sum_{l =1}^{N-1} x^\prime_{l} \sigma_l^z \sigma_{l+1}^z + \hbar \sum_{l=1}^N x^\prime_{N+l-1} \sigma_l^z.
\end{equation}  
This mapping preserves the structure of the classical input in the Hamiltonian couplings and fields, allowing the QA to process the higher-dimensional MNIST data.

\begin{figure}[htbp]
\centering
\includegraphics[width=0.42\textwidth]{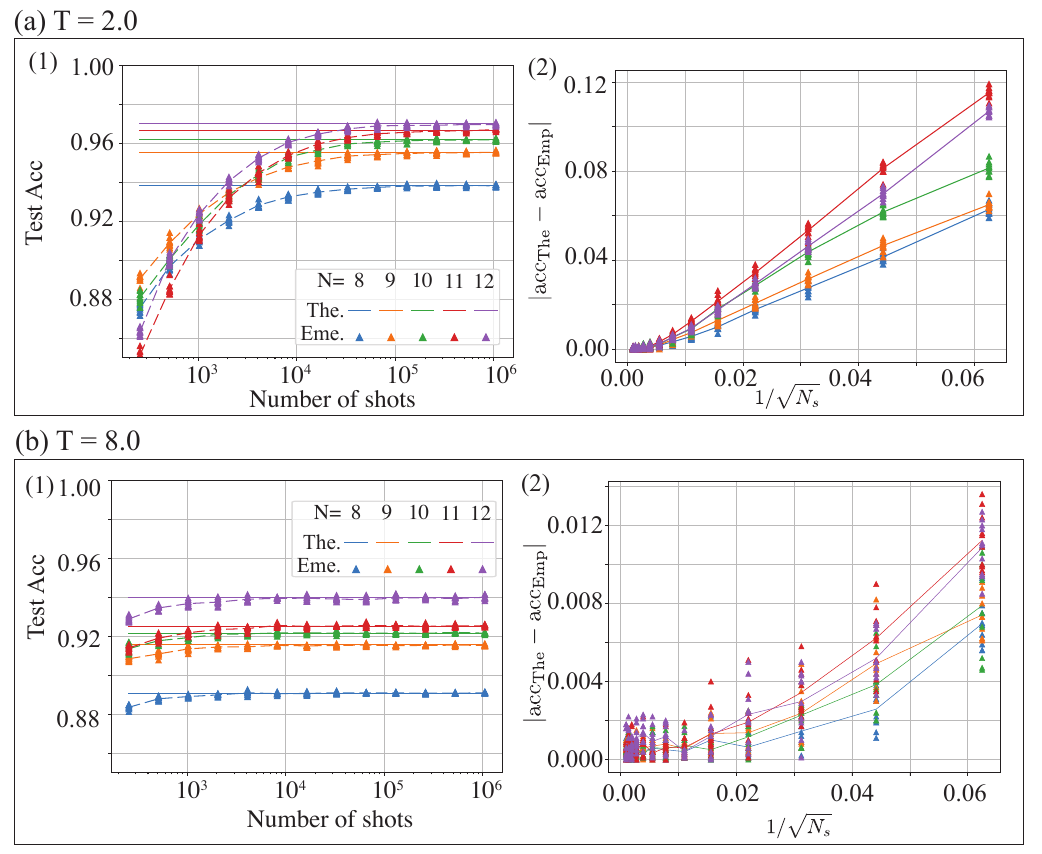}
\caption{
\textbf{Classification performance under finite-shot conditions and associated error.}   Classification results are shown for two annealing times: $T = 2.0$ (a) and $T = 8.0$ (b).  For each case, panel (1) presents the test accuracy using the full theoretical distribution (solid lines), while panel (2) shows the absolute difference between performance based on theoretical versus estimated distributions for various numbers of qubits $N$. In panel (2), the vertical axis corresponds to the performance error, and the horizontal axis is $1/\sqrt{N_s}$. 
}
\label{sampling_T_2_T_8}
\end{figure}

To evaluate the performance of our model on the MNIST dataset, we analyzed the classification accuracy for different numbers of qubits ($N = 8 \sim 12$) and annealing times. High classification accuracy is observed in regions corresponding to short annealing times, while performance gradually degrades as the annealing time increases. By increasing the number of qubits, a higher peak classification accuracy can be achieved, as illustrated in Figs.~\ref{sampling_T_2_T_8} (a-1) and (b-1). Figs.~\ref{sampling_T_2_T_8} (a-2) and (b-2) show the absolute error between the performance using theoretical and reconstructed distributions (vertical axis) versus $1/\sqrt{N_s}$ (horizontal axis). These results indicate that longer annealing times ($T = 8.0$) enable faster convergence to the theoretical performance with fewer shots. This behavior arises from the increased localization in the output probability distribution, which can be quantitatively confirmed through the analysis of various participation ratios.

The primary cost factors for the quantum computation are the execution time of the QA (annealing time) and the number of shots required to reconstruct the output probability distribution.   These factors exhibit a trade-off that is mediated by the averaged participation ratio (APR) over the traning dataset ($\mathbb{D}$),  $\mathrm{APR}(\mathbb{D}) = \left\langle 1/\sum_{l=1}^N | \langle l | \phi \rangle |^4 \right\rangle_{\mathbb{D}}$, where $|l\rangle$ denotes the computational basis (or projection basis), and $|\phi \rangle$ represents the quantum state after annealing.  The APR strongly influences both the learning performance and the number of shots required for reliable sampling~\cite{Akitada2025}.

\begin{figure}[htbp]
\centering
\includegraphics[width=0.42\textwidth]{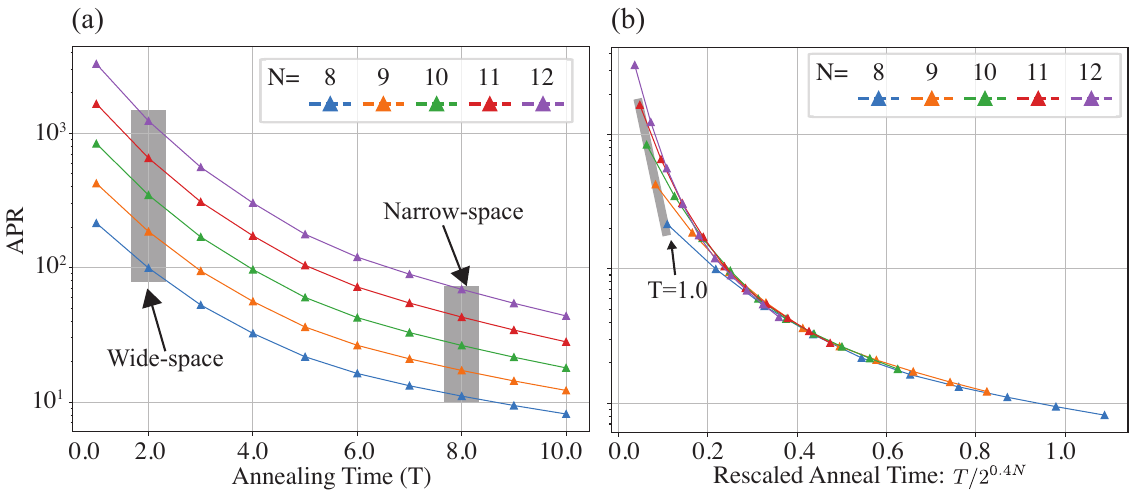}
\caption{
\textbf{Relationship between APR and annealing time for the MNIST dataset.} (a) Direct relationship between APR and annealing time. The positions corresponding to $T=2.0$ and $T=8.0$ are highlighted in gray. (b) Relationship between APR and annealing time rescaled by the number of qubits. The result for $T=1.0$, highlighted in gray, clearly deviates from the rest of the data, indicating that insufficient quantum dynamics occur at such short annealing times.
}
\label{APR_AT_RAT.pdf}
\end{figure}

Fig.~\ref{APR_AT_RAT.pdf} (a) illustrates the connection between annealing time and APR.  For a fixed annealing time, the APR increases exponentially with the number of qubits and is evenly spaced on a logarithmic scale. With a constant APR, the number of required shots may also grow exponentially as the number of qubits increases.   Interestingly, as the annealing time increases, the gap between APR values for different numbers of qubits tends to shrink, indicating that the exponential growth rate of the APR may slow down at longer annealing times.

Fig.~\ref{APR_AT_RAT.pdf} (b) presents APR as a function of the rescaled annealing time per qubit. The figure shows that the relationship between APR and rescaled annealing time ($T/2^{0.4N}$) converges toward a universal curve. This suggests that the annealing time required to reach a target APR may increase exponentially with the number of qubits. To achieve a probability distribution with the same magnitude of APR and maintain a comparable number of required shots across different qubit counts, the compression time must increase exponentially with qubit number. 

To conclude, the computational cost of our model rises exponentially with the number of qubits. However, there is no clear evidence that the number of shots required for the test performance to converge exhibits exponential growth. Although further research is needed to fully understand the requirements and underlying mechanisms of QML with QAs, our experimental demonstration confirms the feasibility of the model, and highlights its potential for more practical classification tasks.

\begin{acknowledgments}
We thank W. J. Munro for the useful discussions.  This work was partly funded by MEXT Quantum Leap Flagship Program (MEXT Q-LEAP) JPMXS0118069605, the JSPS KAKENHI grant no. 21H04880, COI-NEXT JPMJPF2221, and  JSPS KAKENHI under Grant No. 25K21306. 
\end{acknowledgments}

\bibliography{Mybib}

@article{Kadowaki1998,
	author = {Kadowaki, Tadashi and Nishimori, Hidetoshi},
	doi = {10.1103/PhysRevE.58.5355},
	issue = {5},
	journal = {Phys. Rev. E},
	month = {Nov},
	numpages = {0},
	pages = {5355--5363},
	publisher = {American Physical Society},
	title = {Quantum annealing in the transverse Ising model},
	url = {https://link.aps.org/doi/10.1103/PhysRevE.58.5355},
	volume = {58},
	year = {1998}}

@article{Noori2020,
	author = {Noori, Moslem and Vedaie, Seyed Shakib and Singh, Inderpreet and Crawford, Daniel and Oberoi, Jaspreet S. and Sanders, Barry C. and Zahedinejad, Ehsan},
	doi = {10.1103/PhysRevApplied.14.034034},
	issue = {3},
	journal = {Phys. Rev. Appl.},
	month = {Sep},
	numpages = {10},
	pages = {034034},
	publisher = {American Physical Society},
	title = {Analog-Quantum Feature Mapping for Machine-Learning Applications},
	url = {https://link.aps.org/doi/10.1103/PhysRevApplied.14.034034},
	volume = {14},
	year = {2020}}

@article{Edwards1972aa,
	author = {J T Edwards and D J Thouless},
	date = {1972/04/27},
	doi = {10.1088/0022-3719/5/8/007},
	isbn = {0022-3719},
	journal = {Journal of Physics C: Solid State Physics},
	number = {8},
	pages = {807},
	title = {Numerical studies of localization in disordered systems},
	url = {https://doi.org/10.1088/0022-3719/5/8/007},
	volume = {5},
	year = {1972}}

@article{Mujal2023,
	author = {Mujal, Pere and Mart{\'\i}nez-Pe{\~{n}}a, Rodrigo and Giorgi, Gian Luca and Soriano, Miguel C. and Zambrini, Roberta},
	day = {23},
	doi = {10.1038/s41534-023-00682-z},
	issn = {2056-6387},
	journal = {npj Quantum Information},
	month = {Feb},
	number = {1},
	pages = {16},
	title = {Time-series quantum reservoir computing with weak and projective measurements},
	url = {https://doi.org/10.1038/s41534-023-00682-z},
	volume = {9},
	year = {2023}}

@article{Pfeffer2022,
	author = {Pfeffer, Philipp and Heyder, Florian and Schumacher, J\"org},
	doi = {10.1103/PhysRevResearch.4.033176},
	issue = {3},
	journal = {Phys. Rev. Res.},
	month = {Sep},
	numpages = {14},
	pages = {033176},
	publisher = {American Physical Society},
	title = {Hybrid quantum-classical reservoir computing of thermal convection flow},
	url = {https://link.aps.org/doi/10.1103/PhysRevResearch.4.033176},
	volume = {4},
	year = {2022}}

@article{Bravo2022,
	author = {Bravo, Rodrigo Araiza and Najafi, Khadijeh and Gao, Xun and Yelin, Susanne F.},
	doi = {10.1103/PRXQuantum.3.030325},
	issue = {3},
	journal = {PRX Quantum},
	month = {Aug},
	numpages = {19},
	pages = {030325},
	publisher = {American Physical Society},
	title = {Quantum Reservoir Computing Using Arrays of Rydberg Atoms},
	url = {https://link.aps.org/doi/10.1103/PRXQuantum.3.030325},
	volume = {3},
	year = {2022}}

@article{Suzuki2022,
	author = {Suzuki, Yudai and Gao, Qi and Pradel, Ken C. and Yasuoka, Kenji and Yamamoto, Naoki},
	day = {25},
	doi = {10.1038/s41598-022-05061-w},
	issn = {2045-2322},
	journal = {Scientific Reports},
	month = {Jan},
	number = {1},
	pages = {1353},
	title = {Natural quantum reservoir computing for temporal information processing},
	url = {https://doi.org/10.1038/s41598-022-05061-w},
	volume = {12},
	year = {2022}}

@article{Nakajima2019,
	author = {Nakajima, Kohei and Fujii, Keisuke and Negoro, Makoto and Mitarai, Kosuke and Kitagawa, Masahiro},
	doi = {10.1103/PhysRevApplied.11.034021},
	issue = {3},
	journal = {Phys. Rev. Applied},
	month = {Mar},
	numpages = {17},
	pages = {034021},
	publisher = {American Physical Society},
	title = {Boosting Computational Power through Spatial Multiplexing in Quantum Reservoir Computing},
	url = {https://link.aps.org/doi/10.1103/PhysRevApplied.11.034021},
	volume = {11},
	year = {2019}}

@article{Mitarai2018,
	author = {Mitarai, K. and Negoro, M. and Kitagawa, M. and Fujii, K.},
	doi = {10.1103/PhysRevA.98.032309},
	issue = {3},
	journal = {Phys. Rev. A},
	month = {Sep},
	numpages = {6},
	pages = {032309},
	publisher = {American Physical Society},
	title = {Quantum circuit learning},
	url = {https://link.aps.org/doi/10.1103/PhysRevA.98.032309},
	volume = {98},
	year = {2018}}

@misc{fujii2020,
	archiveprefix = {arXiv},
	author = {Keisuke Fujii and Kohei Nakajima},
	eprint = {2011.04890},
	primaryclass = {quant-ph},
	title = {Quantum reservoir computing: a reservoir approach toward quantum machine learning on near-term quantum devices},
	year = {2020}}

@article{Fujii2018,
	author = {Fujii, Keisuke and Kobayashi, Hirotada and Morimae, Tomoyuki and Nishimura, Harumichi and Tamate, Shuhei and Tani, Seiichiro},
	doi = {10.1103/PhysRevLett.120.200502},
	issue = {20},
	journal = {Phys. Rev. Lett.},
	month = {May},
	numpages = {6},
	pages = {200502},
	publisher = {American Physical Society},
	title = {Impossibility of Classically Simulating One-Clean-Qubit Model with Multiplicative Error},
	url = {https://link.aps.org/doi/10.1103/PhysRevLett.120.200502},
	volume = {120},
	year = {2018}}

@article{Fujii2017,
	author = {Fujii, Keisuke and Nakajima, Kohei},
	doi = {10.1103/PhysRevApplied.8.024030},
	issue = {2},
	journal = {Phys. Rev. Appl.},
	month = {Aug},
	numpages = {20},
	pages = {024030},
	publisher = {American Physical Society},
	title = {Harnessing Disordered-Ensemble Quantum Dynamics for Machine Learning},
	url = {https://link.aps.org/doi/10.1103/PhysRevApplied.8.024030},
	volume = {8},
	year = {2017}}

@article{Hayashi2023,
	author = {Hayashi, Aoi and Sakurai, Akitada and Nishio, Shin and Munro, William J. and Nemoto, Kae},
	doi = {10.1103/PhysRevA.108.042609},
	issue = {4},
	journal = {Phys. Rev. A},
	month = {Oct},
	numpages = {11},
	pages = {042609},
	publisher = {American Physical Society},
	title = {Impact of the form of weighted networks on the quantum extreme reservoir computation},
	url = {https://link.aps.org/doi/10.1103/PhysRevA.108.042609},
	volume = {108},
	year = {2023}}

@article{mehta2025,
	author = {Mehta, Vrinda and De Raedt, Hans and Michielsen, Kristel and Jin, Fengping},
	doi = {10.1103/3yk8-qn64},
	issue = {3},
	journal = {Phys. Rev. A},
	month = {Sep},
	numpages = {14},
	pages = {032616},
	publisher = {American Physical Society},
	title = {Understanding the physics of D-Wave annealers: From Schr\"odinger to Lindblad to Markovian dynamics},
	url = {https://link.aps.org/doi/10.1103/3yk8-qn64},
	volume = {112},
	year = {2025}}

@misc{boothby2020,
	archiveprefix = {arXiv},
	author = {Kelly Boothby and Paul Bunyk and Jack Raymond and Aidan Roy},
	eprint = {2003.00133},
	primaryclass = {quant-ph},
	title = {Next-Generation Topology of D-Wave Quantum Processors},
	url = {https://arxiv.org/abs/2003.00133},
	year = {2020}}

@article{Akitada2022,
	author = {Sakurai, Akitada and Estarellas, Marta P. and Munro, William J. and Nemoto, Kae},
	doi = {10.1103/PhysRevApplied.17.064044},
	issue = {6},
	journal = {Phys. Rev. Appl.},
	month = {Jun},
	numpages = {10},
	pages = {064044},
	publisher = {American Physical Society},
	title = {Quantum Extreme Reservoir Computation Utilizing Scale-Free Networks},
	url = {https://link.aps.org/doi/10.1103/PhysRevApplied.17.064044},
	volume = {17},
	year = {2022}}

@misc{haykin2009,
	author = {Haykin, Simon},
	publisher = {Pearson Education India},
	title = {Neural networks and learning machines, 3/E},
	year = {2009}}

@article{scikit-learn,
	author = {Pedregosa, F. and Varoquaux, G. and Gramfort, A. and Michel, V. and Thirion, B. and Grisel, O. and Blondel, M. and Prettenhofer, P. and Weiss, R. and Dubourg, V. and Vanderplas, J. and Passos, A. and Cournapeau, D. and Brucher, M. and Perrot, M. and Duchesnay, E.},
	journal = {Journal of Machine Learning Research},
	pages = {2825--2830},
	title = {Scikit-learn: Machine Learning in {P}ython},
	volume = {12},
	year = {2011}}

@unpublished{Wurtz:2022aa,
	author = {Wurtz, J and Lopes, P and Gemelke, N and Keesling, A and Wang, S},
	title = {Industry applications of neutral-atom quantum computing solving independent set problems},
	year = {2022}}

@article{Crawford2018,
	address = {Paramus, NJ},
	author = {Crawford, Daniel and Levit, Anna and Ghadermarzy, Navid and Oberoi, Jaspreet S. and Ronagh, Pooya},
	issn = {1533-7146},
	issue_date = {February 2018},
	journal = {Quantum Info. Comput.},
	month = feb,
	number = {1--2},
	numpages = {24},
	pages = {51--74},
	publisher = {Rinton Press, Incorporated},
	title = {Reinforcement learning using quantum boltzmann machines},
	volume = {18},
	year = {2018}}

@misc{neven2009,
	archiveprefix = {arXiv},
	author = {Hartmut Neven and Vasil S. Denchev and Geordie Rose and William G. Macready},
	eprint = {0912.0779},
	primaryclass = {quant-ph},
	title = {Training a Large Scale Classifier with the Quantum Adiabatic Algorithm},
	url = {https://arxiv.org/abs/0912.0779},
	year = {2009}}

@article{Mott2017aa,
	author = {Mott, Alex and Job, Joshua and Vlimant, Jean-Roch and Lidar, Daniel and Spiropulu, Maria},
	doi = {10.1038/nature24047},
	journal = {Nature},
	month = {Oct},
	number = {7676},
	pages = {375--379},
	title = {Solving a Higgs optimization problem with quantum annealing for machine learning.},
	volume = {550},
	year = {2017}}

@misc{mcgeoch2022adiabatic,
	author = {McGeoch, Catherine C},
	publisher = {Springer Nature},
	title = {Adiabatic quantum computation and quantum annealing: Theory and practice},
	year = {2022}}

@article{Li2018,
	author = {Li, Richard Y. and Di Felice, Rosa and Rohs, Remo and Lidar, Daniel A.},
	date = {2018/02/21},
	doi = {10.1038/s41534-018-0060-8},
	id = {Li2018},
	isbn = {2056-6387},
	journal = {npj Quantum Information},
	number = {1},
	pages = {14},
	title = {Quantum annealing versus classical machine learning applied to a simplified computational biology problem},
	url = {https://doi.org/10.1038/s41534-018-0060-8},
	volume = {4},
	year = {2018}}

@article{Salloum2025,
	author = {Salloum, Hadi and Sabbagh, Kamil and Savchuk, Vladislav and Lukin, Ruslan and Orabi, Osama and Isangulov, Marat and Mazzara, Manuel},
	doi = {10.1109/ACCESS.2025.3531391},
	journal = {IEEE Access},
	pages = {16263-16287},
	title = {Performance of Quantum Annealing Machine Learning Classification Models on ADMET Datasets},
	volume = {13},
	year = {2025}}

@article{Sakurai25Opt,
	author = {Akitada Sakurai and Aoi Hayashi and William John Munro and Kae Nemoto},
	doi = {10.1364/OPTICAQ.541432},
	journal = {Optica Quantum},
	month = {Jun},
	number = {3},
	pages = {238--245},
	publisher = {Optica Publishing Group},
	title = {Quantum optical reservoir computing powered by boson sampling},
	url = {https://opg.optica.org/opticaq/abstract.cfm?URI=opticaq-3-3-238},
	volume = {3},
	year = {2025}}

@article{Akitada2025,
	author = {Sakurai, Akitada and Hayashi, Aoi and Munro, W. J. and Nemoto, Kae},
	doi = {10.1103/PhysRevA.111.052432},
	issue = {5},
	journal = {Phys. Rev. A},
	month = {May},
	numpages = {16},
	pages = {052432},
	publisher = {American Physical Society},
	title = {Simple Hamiltonian dynamics as a powerful resource for image classification},
	url = {https://link.aps.org/doi/10.1103/PhysRevA.111.052432},
	volume = {111},
	year = {2025}}

@article{Mach2011,
	author = {Duchi, John and Hazan, Elad and Singer, Yoram},
	issn = {1532-4435},
	issue_date = {2/1/2011},
	journal = {J. Mach. Learn. Res.},
	month = jul,
	number = {null},
	numpages = {39},
	pages = {2121--2159},
	publisher = {JMLR.org},
	title = {Adaptive Subgradient Methods for Online Learning and Stochastic Optimization},
	volume = {12},
	year = {2011}}

@article{Liu2018,
	article-number = {380},
	author = {Liu, Jeremy and Spedalieri, Federico M. and Yao, Ke-Thia and Potok, Thomas E. and Schuman, Catherine and Young, Steven and Patton, Robert and Rose, Garrett S. and Chamka, Gangotree},
	doi = {10.3390/e20050380},
	issn = {1099-4300},
	journal = {Entropy},
	number = {5},
	pubmedid = {33265470},
	title = {Adiabatic Quantum Computation Applied to Deep Learning Networks},
	url = {https://www.mdpi.com/1099-4300/20/5/380},
	volume = {20},
	year = {2018}}

@article{Deng2012,
	author = {Deng, Li},
	doi = {10.1109/MSP.2012.2211477},
	journal = {IEEE Signal Processing Magazine},
	number = {6},
	pages = {141-142},
	title = {The MNIST Database of Handwritten Digit Images for Machine Learning Research [Best of the Web]},
	volume = {29},
	year = {2012}}

@misc{DIGIT1998,
	author = {Alpaydin, E. and Kaynak, C.},
	howpublished = {UCI Machine Learning Repository},
	note = {{DOI}: https://doi.org/10.24432/C50P49},
	title = {{Optical Recognition of Handwritten Digits}},
	year = {1998}}

@article{King2025,
	author = {King, Andrew D. and Nocera, Alberto and Rams, Marek M. and Dziarmaga, Jacek and Wiersema, Roeland and Bernoudy, William and Raymond, Jack and Kaushal, Nitin and Heinsdorf, Niclas and Harris, Richard and Boothby, Kelly and Altomare, Fabio and Asad, Mohsen and Berkley, Andrew J. and Boschnak, Martin and Chern, Kevin and Christiani, Holly and Cibere, Samantha and Connor, Jake and Dehn, Martin H. and Deshpande, Rahul and Ejtemaee, Sara and Farre, Pau and Hamer, Kelsey and Hoskinson, Emile and Huang, Shuiyuan and Johnson, Mark W. and Kortas, Samuel and Ladizinsky, Eric and Lanting, Trevor and Lai, Tony and Li, Ryan and MacDonald, Allison J. R. and Marsden, Gaelen and McGeoch, Catherine C. and Molavi, Reza and Oh, Travis and Neufeld, Richard and Norouzpour, Mana and Pasvolsky, Joel and Poitras, Patrick and Poulin-Lamarre, Gabriel and Prescott, Thomas and Reis, Mauricio and Rich, Chris and Samani, Mohammad and Sheldan, Benjamin and Smirnov, Anatoly and Sterpka, Edward and Trullas Clavera, Berta and Tsai, Nicholas and Volkmann, Mark and Whiticar, Alexander M. and Whittaker, Jed D. and Wilkinson, Warren and Yao, Jason and Yi, T. J. and Sandvik, Anders W. and Alvarez, Gonzalo and Melko, Roger G. and Carrasquilla, Juan and Franz, Marcel and Amin, Mohammad H.},
	date = {2025/04/11},
	doi = {10.1126/science.ado6285},
	journal = {Science},
	journal1 = {Science},
	journal2 = {Science},
	month = {2025/12/07},
	n2 = {Quantum computers hold the promise of solving certain problems that lie beyond the reach of conventional computers. However, establishing this capability, especially for impactful and meaningful problems, remains a central challenge. Here, we show that superconducting quantum annealing processors can rapidly generate samples in close agreement with solutions of the Schr{\"o}dinger equation. We demonstrate area-law scaling of entanglement in the model quench dynamics of two-, three-, and infinite-dimensional spin glasses, supporting the observed stretched-exponential scaling of effort for matrix-product-state approaches. We show that several leading approximate methods based on tensor networks and neural networks cannot achieve the same accuracy as the quantum annealer within a reasonable time frame. Thus, quantum annealers can answer questions of practical importance that may remain out of reach for classical computation. Quantum computers should be able to solve certain problems that classical computers cannot; however, at the current stage of development, imperfections in quantum computing hardware diminish this comparative advantage. King et al. contrasted the performance of their quantum annealing processor to state-of-the-art classical simulations of topical problems such as the quantum dynamics of the transverse-field Ising model. The researchers found that across a range of graph topologies, the quantum processor was able to outperform classical simulations. The results provide a challenge to classical computing, in which method improvement has in the past tempered claims of quantum advantage. ?Jelena Stajic},
	number = {6743},
	pages = {199--204},
	publisher = {American Association for the Advancement of Science},
	title = {Beyond-classical computation in quantum simulation},
	type = {doi: 10.1126/science.ado6285},
	url = {https://doi.org/10.1126/science.ado6285},
	volume = {388},
	year = {2025},
	year1 = {2025}}

@article{King2023,
	author = {King, Andrew D. and Raymond, Jack and Lanting, Trevor and Harris, Richard and Zucca, Alex and Altomare, Fabio and Berkley, Andrew J. and Boothby, Kelly and Ejtemaee, Sara and Enderud, Colin and Hoskinson, Emile and Huang, Shuiyuan and Ladizinsky, Eric and MacDonald, Allison J. R. and Marsden, Gaelen and Molavi, Reza and Oh, Travis and Poulin-Lamarre, Gabriel and Reis, Mauricio and Rich, Chris and Sato, Yuki and Tsai, Nicholas and Volkmann, Mark and Whittaker, Jed D. and Yao, Jason and Sandvik, Anders W. and Amin, Mohammad H.},
	date = {2023/05/01},
	doi = {10.1038/s41586-023-05867-2},
	id = {King2023},
	isbn = {1476-4687},
	journal = {Nature},
	number = {7959},
	pages = {61--66},
	title = {Quantum critical dynamics in a 5,000-qubit programmable spin glass},
	url = {https://doi.org/10.1038/s41586-023-05867-2},
	volume = {617},
	year = {2023}}

@misc{Dwave,
	author = {{D-Wave Systems Inc.}},
	howpublished = {\url{https://docs.dwavesys.com/docs/latest/index.html}},
	note = {Accessed: 2025-12-25},
	title = {{D-Wave Leap Quantum Cloud Service}},
	year = {2025}}

@article{Sato:2021,
	author = {Sato, Takehito and Ohzeki, Masayuki and Tanaka, Kazuyuki},
	date = {2021/06/29},
	doi = {10.1038/s41598-021-92295-9},
	id = {Sato2021},
	isbn = {2045-2322},
	journal = {Scientific Reports},
	number = {1},
	pages = {13523},
	title = {Assessment of image generation by quantum annealer},
	url = {https://doi.org/10.1038/s41598-021-92295-9},
	volume = {11},
	year = {2021}}

@misc{Dwave2025,
	author = {{D-Wave Systems Inc.}},
	howpublished = {\url{https://docs.dwavequantum.com/en/latest/quantum_research/solver_properties_specific.html}},
	title = {{D-Wave Documentation}},
	year = {2025}}

@article{Shibukawa2024,
	author = {Shibukawa, Ryosuke and Tamura, Ryo and Tsuda, Koji},
	doi = {10.1103/PhysRevResearch.6.043050},
	issue = {4},
	journal = {Phys. Rev. Res.},
	month = {Oct},
	numpages = {7},
	pages = {043050},
	publisher = {American Physical Society},
	title = {Boltzmann sampling with quantum annealers via fast Stein correction},
	url = {https://link.aps.org/doi/10.1103/PhysRevResearch.6.043050},
	volume = {6},
	year = {2024}}

@article{Santoro2006,
	author = {Santoro, Giuseppe E and Tosatti, Erio},
	doi = {10.1088/0305-4470/39/36/R01},
	journal = {Journal of Physics A: Mathematical and General},
	month = {aug},
	number = {36},
	pages = {R393},
	title = {Optimization using quantum mechanics: quantum annealing through adiabatic evolution},
	url = {https://dx.doi.org/10.1088/0305-4470/39/36/R01},
	volume = {39},
	year = {2006}}

@article{DasArnab2008,
	author = {Das, Arnab and Chakrabarti, Bikas K.},
	doi = {10.1103/RevModPhys.80.1061},
	issue = {3},
	journal = {Rev. Mod. Phys.},
	month = {Sep},
	numpages = {0},
	pages = {1061--1081},
	publisher = {American Physical Society},
	title = {Colloquium: Quantum annealing and analog quantum computation},
	url = {https://link.aps.org/doi/10.1103/RevModPhys.80.1061},
	volume = {80},
	year = {2008}}

@article{Morita2008,
	author = {Morita, Satoshi and Nishimori, Hidetoshi},
	doi = {10.1063/1.2995837},
	eprint = {https://pubs.aip.org/aip/jmp/article-pdf/doi/10.1063/1.2995837/13869474/125210\_1\_online.pdf},
	issn = {0022-2488},
	journal = {Journal of Mathematical Physics},
	month = {12},
	number = {12},
	pages = {125210},
	title = {Mathematical foundation of quantum annealing},
	url = {https://doi.org/10.1063/1.2995837},
	volume = {49},
	year = {2008}}

@article{Albash2018,
	author = {Albash, Tameem and Lidar, Daniel A.},
	doi = {10.1103/RevModPhys.90.015002},
	issue = {1},
	journal = {Rev. Mod. Phys.},
	month = {Jan},
	numpages = {64},
	pages = {015002},
	publisher = {American Physical Society},
	title = {Adiabatic quantum computation},
	url = {https://link.aps.org/doi/10.1103/RevModPhys.90.015002},
	volume = {90},
	year = {2018}}

@article{Hauke2020,
	author = {Hauke, Philipp and Katzgraber, Helmut G and Lechner, Wolfgang and Nishimori, Hidetoshi and Oliver, William D},
	doi = {10.1088/1361-6633/ab85b8},
	journal = {Reports on Progress in Physics},
	month = {may},
	number = {5},
	pages = {054401},
	publisher = {IOP Publishing},
	title = {Perspectives of quantum annealing: methods and implementations},
	url = {https://dx.doi.org/10.1088/1361-6633/ab85b8},
	volume = {83},
	year = {2020}}

@article{DasArnab2005,
	author = {Das, Arnab and Chakrabarti, Bikas K. and Stinchcombe, Robin B.},
	doi = {10.1103/PhysRevE.72.026701},
	issue = {2},
	journal = {Phys. Rev. E},
	month = {Aug},
	numpages = {4},
	pages = {026701},
	publisher = {American Physical Society},
	title = {Quantum annealing in a kinetically constrained system},
	url = {https://link.aps.org/doi/10.1103/PhysRevE.72.026701},
	volume = {72},
	year = {2005}}
\end{document}